\documentclass[10pt,letterpaper]{article}
\usepackage{geometry}
\usepackage{amsmath,amssymb,slashed,bbm}
\usepackage{setspace}

\csname@addtoreset\endcsname{equation}{section}

\makeatletter
\pdfoutput=1

\usepackage[T1]{fontenc}
\usepackage{ae}
\usepackage{aecompl}
\usepackage{soul}

\usepackage[english]{babel}

\usepackage[sf,bf,small]{titlesec}

\usepackage[small,sf,bf,hang]{caption}

\usepackage[multiple]{footmisc}
\long\def\symbolfootnote[#1]#2{\begingroup%
\def\thefootnote{\fnsymbol{footnote}}\footnote[#1]{#2}\endgroup}

\usepackage{titletoc}
\dottedcontents{section}[6mm]{\smallskip }{6mm}{0pc}
\dottedcontents{subsection}[8mm]{\itshape\small }{6mm}{0pc}
\dottedcontents{subsubsection}[16mm]{\itshape\small }{10mm}{0pc}
\def\tableofcontents{\subsection*{\contentsname}\vspace{-2mm}\@starttoc{toc}}

\usepackage[nosort]{cite}

\usepackage[parsep]{collref}
\def \bea  {\begin{eqnarray}}
\def \eea  {\end{eqnarray}}

\begin{document}

\begin{flushright}
MIFPA-13-21\\
QGaSLAB-13-06
\bigskip\bigskip
\par\end{flushright}

\begin{center}
\textsf{\textbf{\Large The low energy limit of the $AdS_3\times S^3\times M_4$ spinning string \smallskip\smallskip }}\\
\textsf{\textbf{\Large  }}
\par\end{center}{\Large \par}

\begin{singlespace}
\begin{center}
Per Sundin$^{1}$ and  Linus Wulff$^{2}$ \bigskip \\

{\small $^{1}$}\emph{\small{} The Laboratory for Quantum Gravity \& Strings}\\
\emph{\small Department of Mathematics and Applied Mathematics, }\\
\emph{\small University of Cape Town,}\\
\emph{\small Private Bag, Rondebosch, 7700, South Africa}{\small }\\
\emph{\small nidnus.rep@gmail.com}\vspace{0.2cm}

{\small $^{2}$}\emph{\small{} George P. \& Cynthia Woods Mitchell Institute for Fundamental Physics and Astronomy,}\\
\emph{\small Texas A\&M University, College Station, }\\
\emph{\small TX 77843, USA}\\
\emph{\small linus@physics.tamu.edu }{\small \bigskip }\emph{\small }\\
\par\end{center}{\small \par}
\end{singlespace}

%\begin{center}
%Draft printed \bigskip  \today , from file \jobname.
%\par\end{center}
\bigskip\bigskip

\subsection*{\hspace{9mm}Abstract}
\begin{quote}
We derive the low-energy effective action for the spinning (GKP) string in $AdS_3\times S^3\times M_4$ where $M_4=S^3\times S^1$ or $T^4$. In the first case the action consists of two $O(4)$ non-linear sigma models which are coupled through their interaction with four massless Majorana fermions (plus one free decoupled scalar). While in the second case it consists of one $O(4)$ sigma model coupled to four Majorana fermions together with four free scalars from the $T^4$. We show that these models are classically integrable by constructing their Lax connections.
\bigskip
\thispagestyle{empty}
\end{quote}

\newpage
\tableofcontents
\setcounter{page}{1}

\section{Introduction}

String theory on $AdS_3 \times S^3 \times M_4$ with $M_4=(S^3\times S^1,\,T^4)$ preserving 16 supercharges and supported by pure Ramond-Ramond (RR) flux arises as the gravity side of the $AdS_3 / CFT_2$-correspondence.\footnote{For the case of both NSNS and RR flux see \cite{Cagnazzo:2012se}.} As in earlier incarnations of $AdS / CFT$ there are strong hints of integrability, both at the classical and quantum level \cite{Babichenko:2009dk}, although much less is known on the CFT side. Sparked by the discovery of the integrable structure, $AdS_3 / CFT_2$ has recently enjoyed an increased interest in the literature. While strings on $AdS_3\times S^3\times M_4$ have many similarities with strings in $AdS_5 \times S^5$ and $AdS_4\times \mathbbm{CP}^3$ there are nevertheless crucial differences. The biggest issue seems to be the appearance of massless modes on the worldsheet which are difficult to deal with in the standard approach to integrability. Nevertheless, if one focuses on just the massive modes the Bethe equations and the exact S-matrix for these can be derived along similar lines as in earlier $AdS / CFT$ examples, see \cite{Beisert:2010jr} for a review. Based on the algebraic curve technique the first study of the integrable structure was initiated in \cite{Babichenko:2009dk} and later a full set of Bethe equations was conjectured in \cite{OhlssonSax:2011ms} and the exact S-matrix in \cite{Ahn:2012hw}. These conjectures passed a few initial tests but the mixing of modes from the two $S^3$ factors turned out to be problematic \cite{Rughoonauth:2012qd}. This was subsequently addressed in \cite{Borsato:2012ud,Borsato:2012ss} and later confirmed to match with the tree-level worldsheet S-matrix in \cite{Sundin:2013ypa}.\footnote{A mismatch appears at one loop (at least in the $T^4$ limit). This issue was resolved for $AdS_3\times S^3\times T^4$ with the exact S-matrix conjectured in \cite{Borsato:2013qpa}.} As in other integrable $AdS/CFT$ examples the underlying symmetry can only determine the Bethe equations up to overall scalar phase factors. The one-loop strong coupling contribution to the phases was engineered in \cite{David:2010yg,Beccaria:2012kb, Beccaria:2012pm}, see also \cite{Abbott:2012dd,Abbott:2013mpa}, and later generalized to all orders in \cite{Borsato:2013hoa}. For the massless excitations the situation is, as mentioned, more complicated and currently it is not known how to include them in the exact solution. However, by looking at the decompactifying case where one $S^3$ blows up, i.e. the $AdS_3\times S^3\times T^4$ limit, some first steps were taken in \cite{Sax:2012jv}.

In this paper we will take a somewhat different approach to trying to understand the integrable structures of the $AdS_3\times S^3\times M_4$ string. We will consider the spinning or Gubser-Klebanov-Polyakov (GKP) string solution \cite{Gubser:2002tv}. By performing a fluctuation analysis around the solution it is known that the excitations come in both massive and massless \cite{Forini:2012bb,Frolov:2002av} modes. Generally, the full Lagrangian describing all excitations is fairly involved. However, by restricting to the low-energy sector, capturing the dynamics of the massless modes, the theory becomes much simpler \cite{Alday:2007mf, Bykov:2010tv}. Since the full sigma model is believed to be integrable beyond the classical level, the low-energy GKP string should inherit this property. This opens up new avenues for testing the quantum integrability, see for example \cite{Basso:2012bw,Basso:2013pxa,Fioravanti:2013eia} for recent results in $AdS_4 / CFT_3$.

For the simplest case of $AdS_5\times S^5$ the massless modes of the GKP string are described by an $O(6)$ sigma model \cite{Alday:2007mf} while the low-energy dynamics of the $AdS_4\times \mathbbm{CP}^3$ GKP string are captured by a $\mathbbm{CP}^3$ sigma model coupled to a Dirac fermion \cite{Bykov:2010tv}. For the $AdS_3\times S^3\times S^3\times S^1$ case we will demonstrate that the low-energy dynamics of the GKP string is described by two $O(4)$ non-linear sigma models coupled to four Majorana fermions together with one decoupled scalar,\footnote{The $AdS_3\times S^3\times T^4$ case is obtained by taking $\alpha\rightarrow1$ giving a single $O(4)$ sigma model coupled to fermions with four decoupled free scalars coming from the $T^4$.}
\begin{eqnarray}
\label{eq:main-result}
\mathcal L&=&
\frac12e_i{}^{\hat a}e^{i\hat a}+\frac12e_i{}^{a'}e^{ia'}+\frac12\partial_iy\partial^iy
+\frac{i}{4}\mathrm{tr}\left(\bar\Psi\rho^i\partial_i\Psi\right)
-\frac{1}{16}(\varepsilon^{\hat a\hat b\hat c}\omega_i{}^{\hat b\hat c}+2\sqrt{\alpha}\,\delta_{ij}e^{j\hat a})\,\mathrm{tr}\big(\bar\Psi\rho^i\Psi\sigma^{\hat a}\big)
\nonumber\\
&&{}
+\frac{1}{16}(\varepsilon^{a'b'c'}\omega_i{}^{b'c'}+2\sqrt{1-\alpha}\,\delta_{ij}e^{ja'})\mathrm{tr}\big(\bar\Psi\rho^i\sigma^{a'}\Psi\big)\,,
\end{eqnarray}
where primed and hatted indices refer to the first and second $S^3$ factor respectively. Here $e^{\hat a}$ and $e^{a'}$ are the vielbeins of the two $S^3$'s and $\omega^{\hat a\hat b}$, $\omega^{a'b'}$ the corresponding spin connections, while the parameter $0\leq\alpha\leq1$ controls the relative size of the two $S^3$'s. The four Majorana fermions $\Psi_I$ have been grouped into a $2\times2$ matrix $\Psi=\sigma^I\Psi_I$. The Pauli matrices $\sigma^{\hat a}$ and $\sigma^{a'}$ and the trace act in this space while the $\rho^i$ are the 2d gamma-matrices acting on the spinor indices only. See section \ref{sec:2dform} for more details. Note that the presence of $\delta_{ij}$ in the coupling to the fermions explicitly breaks the 2d Lorentz invariance. This is a novel property as compared to the $AdS_5\times S^5$ and $AdS_4\times \mathbbm{CP}^3$ strings whose low-energy dynamics are described by relativistic sigma models. Note also the absence of $\Psi^4$-terms which were present in the $AdS_4\times \mathbbm{CP}^3$ case. Despite these differences we will show that this model is also integrable, at least at the classical level.

The structure of the paper is as follows. We begin by describing the structure of the $AdS_3\times S^3\times S^3\times S^1$ Green-Schwarz string action up to fourth order in fermions in section \ref{sec:action}. We then introduce suitable coordinates and derive the low-energy effective action in sections \ref{sec:parameterization} and \ref{sec:low-energy}. This is done by first integrating out the massive bosonic coordinate and then putting all remaining massive fields to zero by hand. We also introduce a kappa symmetry gauge-fixing which turns out to be useful. It is shown that the low-energy effective action reduces to (\ref{eq:main-result}). In section \ref{sec:integrability} we show that this model is classically integrable by constructing its Lax representation. We end the paper with some conclusions. 

\section{The Green-Schwarz string in $AdS_3\times S^3\times M_4$}
\label{sec:action}
The type II Green-Schwarz superstring action in a general supergravity background can be expanded in the fermions as
\begin{equation}
\label{action}
S=-T\int d^2\sigma\,\mathcal L\,,\qquad \mathcal L=\mathcal L^{(0)}+\mathcal L^{(2)}+\mathcal L^{(4)}+\ldots
\end{equation}
The purely bosonic Lagrangian is
\begin{equation}
\label{eq:L0}
\mathcal L^{(0)}=\frac{1}{2}\gamma^{ij}e_i{}^Ae_j{}^B\eta_{AB}\,,\qquad\gamma^{ij}=\sqrt{-g}\,g^{ij}\,,
\end{equation}
where $e_i{}^A(X)$ $(A=0,1,\cdots,9)$ are the vielbeins of the purely bosonic part of the background pulled back to the worldsheet and $g_{ij}$ is an independent worldsheet metric with $g=\det{g_{ij}}$.

The terms quadratic in fermions take the form \cite{Cvetic:1999zs}
\begin{equation}
\label{eq:L2}
\mathcal L^{(2)}=ie_i{}^A\,\bar\Theta\Gamma_AK^{ij}{\mathcal D}_j\Theta\,,\qquad K^{ij}=\gamma^{ij}-\varepsilon^{ij}\Gamma_{11}\,.
\end{equation}
The appearance of the matrix $K^{ij}$ is related to kappa symmetry. The Killing spinor derivative $\mathcal D$ is given below.

The quartic fermion terms in the action were recently found in \cite{Wulff:2013kga}. They take the form\footnote{Our normalization of $\Theta$ differs from that of \cite{Wulff:2013kga} by a factor of $\sqrt2$.}
\bea
\label{eq:L4}
\mathcal L^{(4)}&=&
-\frac{1}{2}\bar\Theta\Gamma^A\mathcal{D}_i\Theta\,\bar\Theta\Gamma_A K^{ij}\mathcal{D}_j\Theta 
+\frac{i}{6}e_i{}^A\,\bar\Theta\Gamma_AK^{ij}\mathcal M\mathcal D_j\Theta 
+\frac{i}{48}e_i{}^Ae_j{}^B\,\bar\Theta\Gamma_AK^{ij}\big(M+\tilde M\big)S\Gamma_B\Theta
\nonumber\\
&&{}
+\frac{1}{48}e_i{}^Ae_j{}^B\,\bar\Theta\Gamma_A{}^{CD}K^{ij}\Theta\,\big(3\bar\Theta\Gamma_BU_{CD}\Theta-2\bar\Theta\Gamma_CU_{DB}\Theta\big) 
\nonumber\\ 
&&{} 
-\frac{1}{48}e_i{}^Ae_j{}^B\,\bar\Theta\Gamma_A{}^{CD}\Gamma_{11}K^{ij}\Theta\,\big(3\bar\Theta\Gamma_B\Gamma_{11}U_{CD}\Theta+2\bar\Theta\Gamma_C\Gamma_{11}U_{DB}\Theta\big)\,.
\eea
The definition of $\mathcal D$, $\mathcal M$, $M$ and $U_{AB}$ for a general type II supergravity background can be found in \cite{Wulff:2013kga}. Here we will only give the expressions for the case of interest here: the type IIA $AdS_3\times S^3\times S^3\times S^1$ supergravity solution with RR four-form flux. $\Theta$ is taken to be a 32-component Majorana spinor and the Killing spinor derivative is given by
\begin{equation}
\label{E}
{\mathcal D_i}\Theta=(\partial_i-\frac{1}{4}\omega_i{}^{AB}\Gamma_{AB}+\frac{1}{8}e_i{}^A\,S\Gamma_A)\Theta\quad\mbox{where}\qquad S=-4\Gamma^{0129}\big(1-\mathcal P\big)\,.
\end{equation}
Here $\mathcal P$ is a projection matrix given by
\begin{equation}
\label{calP}
\mathcal P=\frac{1}{2}(1+\sqrt\alpha\,\Gamma^{012345}+\sqrt{1-\alpha}\,\Gamma^{012678})
\end{equation}
and is in fact the projector which singles out the 16 supersymmetries preserved by the background. The $AdS_3$-directions are indexed by $(0,1,2)$ the first $S^3$ by $(3,4,5)$, the second $S^3$ by $(6,7,8)$ and the $S^1$ by $(9)$. The parameter $0\leq\alpha\leq1$ determines the relative size of the two $S^3$'s. In units of the $AdS_3$-radius the $S^3$ radii are
\begin{equation}
\label{eq:radii}
\hat R=\frac{1}{\sqrt\alpha}\,,\qquad R'=\frac{1}{\sqrt{1-\alpha}}\,.
\end{equation}
The case $\alpha=0,1$ corresponds to $AdS_3\times S^3\times T^4$ where one of the three-spheres is decompactified. The remaining objects appearing in (\ref{eq:L4}) reduce, in $AdS_3\times S^3\times S^3\times S^1$, to
\bea
\label{eq:quartic-matrices}
&&{}U_{AB}=\frac{1}{32}S\Gamma_{[A}S\Gamma_{B]}-\frac{1}{4}R_{AB}{}^{CD}\Gamma_{CD}\,,
\nonumber\\
&&{}M^\alpha{}_\beta=
\frac{i}{16}\bar\Theta S\Theta\,\delta^\alpha_\beta 
-\frac{i}{8}\Theta^\alpha\,(\bar\Theta S)_\beta
+\frac{i}{8}(\Gamma^AS\Theta)^\alpha\,(\bar\Theta\Gamma_A)_\beta\,,\qquad \Tilde M=\Gamma_{11}M\Gamma_{11}\,,
\nonumber\\
&&{}\mathcal M^\alpha{}_\beta=M^\alpha{}_\beta +\tilde M^\alpha{}_\beta
+\frac{i}{8}(S\Gamma^A\Theta)^\alpha\,(\bar\Theta\Gamma_A)_\beta
-\frac{i}{16}(\Gamma^{AB}\Theta)^\alpha\,(\bar\Theta\Gamma_AS\Gamma_B)_\beta\,,
\eea
where the nonzero components of the Riemann tensor of $AdS_3\times S^3\times S^3\times S^1$ are
\begin{equation}
\label{eq:Riemann}
R_{ab}{}^{cd}=2\delta_{[a}^c\delta^d_{b]}\,,\qquad R_{\hat a\hat b}{}^{\hat c\hat d}=-2\alpha\delta_{[\hat a}^{\hat c}\delta^{\hat d}_{\hat b]}\,,
\qquad R_{a'b'}{}^{c'd'}=-2(1-\alpha)\delta_{[a'}^{c'}\delta^{d'}_{b']}\,,
\end{equation}
where $a, \hat a$ and $a'$ refer to $AdS_3$, the first $S^3$ and the second $S^3$ respectively.

\section{Parameterization}
\label{sec:parameterization}
The metric of $AdS_3$ in global coordinates is
\begin{equation}
ds^2=-\cosh^2\rho\,dt^2+d\rho^2+\sinh^2\rho\,d\varphi^2\,.
\end{equation}
And the long spinning string solution is given by \cite{Gubser:2002tv}	
\begin{equation}
t=\varphi=\kappa\tau\qquad\rho=\kappa\sigma\,,
\end{equation}
with $\kappa$ a constant. It will be more convenient for our purposes however to use different coordinates. Defining new coordinates by
\begin{eqnarray}
\sinh2\zeta&=&-\sin(t-\varphi)\sinh2\rho\nonumber\\
e^{4iu}&=&e^{2i(t+\varphi)}\frac{\cos(t-\varphi)+i\cosh2\rho\sin(t-\varphi)}{\cos(t-\varphi)-i\cosh2\rho\sin(t-\varphi)}\nonumber\\
\sinh2\chi&=&\frac{\cos(t-\varphi)\sinh2\rho}{\sqrt{1+\sin^2(t-\varphi)\sinh^22\rho}}
\end{eqnarray}
the metric takes the form \cite{Alday:2007mf}
\begin{equation}
ds^2=-du^2+d\chi^2-2\sinh2\zeta\,dud\chi+d\zeta^2\,.
\end{equation}
Defining further
\begin{equation}
u=z^++z^-\,,\qquad\chi=z^+-z^-\,,\qquad\zeta=\log\frac{1+\frac{1}{2}z}{1-\frac{1}{2}z}
\end{equation}
the metric becomes
\bea
\label{eq:z-coordinates}
ds^2=-4dz^+dz^-+4\frac{z+\frac{1}{4}z^3}{\big(1-\frac{1}{4}z^2\big)^2}\Big(-(dz^+)^2+(dz^-)^2\Big)+\frac{dz^2}{\big(1-\frac{1}{4}z^2\big)^2}\,.
\eea
The upshot is that this metric is invariant under constant shifts of the two light-cone coordinates $z^\pm$, something which will prove convenient in solving the Virasoro constraints.

In terms of these new coordinates, the long spinning (GKP) string solution is given by
\bea
\label{eq:GKP}
z^\pm=\kappa\sigma^\pm\,,\qquad \sigma^\pm=\frac{1}{2}(\tau\pm\sigma)\,.
\eea
We will leave the $S^3$ metric unspecified and work directly in terms of the spin connection and vielbeins.

\section{Low-energy effective action}
\label{sec:low-energy}
Having defined the $AdS_3$ metric it is now straightforward to expand the action around the spinning string solution (\ref{eq:GKP}). To find the spectrum we consider the quadratic action. Taking the conformal gauge $\gamma^{ij}=\eta^{ij}$ with $\eta_{+-}=2$ and using (\ref{eq:z-coordinates}) the bosonic action (\ref{eq:L0}) reduces to
\begin{equation}
\mathcal L_2^{(0)}=-2\partial_-z^+\partial_+z^-+2\kappa z(\partial_+z^--\partial_-z^+)
+\frac12\partial_+z\partial_-z+\frac12\partial_+y^{\hat m}\partial_-y_{\hat m}+\frac12\partial_+y^{m'}\partial_-y_{m'}+\frac12\partial_+y\partial_-y\,.
\end{equation}
As we will see below solving the Virasoro constraints will eliminate $z^\pm$ from the physical spectrum and generate a mass term for the remaining $AdS$-coordinate $z$. The seven remaining scalars, three ($y^{\hat m}$) from the first $S^3$, three ($y^{m'}$) from the second $S^3$ and one ($y$) from the $S^1$ remain massless. To find the low-energy effective action we will integrate out the massive boson $z$ leaving only the seven massless bosons.

Let us now turn to the spectrum of the fermions. Using the spinning string solution (\ref{eq:GKP}) in the quadratic fermion action (\ref{eq:L2}) and using (\ref{E}) we find at the quadratic level\footnote{We have rescaled the fermions by a factor $\kappa^{-1/2}$.}
\begin{equation}
\label{eq:L2F}
\mathcal L_2^{(2)}=i\bar\upsilon\Gamma_+P\partial_-\upsilon+i\bar\upsilon\Gamma_-P\partial_+\upsilon
+i\bar\vartheta\Gamma_+P\partial_-\vartheta+i\bar\vartheta\Gamma_-P\partial_+\vartheta
-i\kappa\bar\vartheta\Gamma_{29+-}P\vartheta\,,
\end{equation}
where $\Gamma_\pm=\Gamma_0\pm\Gamma_1$ and 
\begin{equation}
\label{eq:kappaP}
P=\frac12\big(1+\Gamma^{01}\Gamma_{11}\big)
\end{equation}
is the kappa symmetry projection matrix which ensures that only 16 of the 32 components of $\Theta$ are physical. We have used the projection operator $\mathcal P$ defined in (\ref{calP}) to split the fermions into $16+16$
\begin{equation}
\Theta=\mathcal P\Theta+(1-\mathcal P)\Theta=\vartheta+\upsilon\,.
\end{equation}
The 16 fermions $\vartheta$ are in one-to-one correspondence with the supersymmetries of the background. They are the fermions described by the supercoset $\frac{D(2,1;\alpha)\times D(2,1;\alpha)}{SU(1,1)\times SU(2)\times SU(2)}$ \cite{Babichenko:2009dk,Rughoonauth:2012qd} and we refer to them as supercoset fermions. As can be seen from (\ref{eq:L2F}) it is precisely these fermions which acquire mass for the spinning string and since we are interested only in the low-energy effective action we will set them to zero in the following. The 16 non-coset fermions $\upsilon$ are massless and they are to be kept in the low-energy effective action. We see that it is important that we started with the full Green-Schwarz superstring action. If we had tried to use instead a partially kappa gauge-fixed version like the supercoset action we would have missed these fermions.\footnote{This was true also for the $AdS_4\times \mathbbm{CP}^3$ string, see Bykov \cite{Bykov:2010tv}.}

\begin{table}[ht]
\begin{center}
\begin{tabular}{c|c|c}
Coordinate & Mass & Multiplicity \\ \hline 
$z$ & $2\kappa$ & $1$ \\ \hline 
$y^{\hat m}$, $y^{m'}$, $y$ & $0$ & $7$ \\ \hline
$\upsilon$ & $0$ & $4$\\ \hline 
$\vartheta$ & $\kappa$  & $4$
\end{tabular}
\end{center}
\caption{Spectrum of excitations around the GKP string.}
\label{spectrum}
\end{table}

The spectrum is summarized in table \ref{spectrum}. To get the low-energy effective action we can simply set the massive fermions to zero but the massive boson $z$ should be integrated out more carefully. We will now describe how to do this.

\subsection{Integrating out the massive boson $z$}
Since the discussion here will affect only the the $AdS$-coordinates ($z^\pm,z$) we will work only with the terms in the Lagrangian involving these fields.
For the low-energy effective action only the terms of mass-dimension two or less are relevant. Using the fact that $z^\pm$ have dimension zero, $\upsilon$ has dimension $\frac12$ and $z$ as it will be integrated out effectively has dimension $1$ we get by expanding (\ref{eq:L0}) and (\ref{eq:L2})
\bea
\mathcal{L}_z=
-2\partial_+ z^- \partial_- z^+
+2\kappa\big(\partial_+z^--\partial_- z^+\big)z
+\frac{i}{2}(\partial_+z^-+\partial_-z^+)\,\bar\upsilon\Gamma_{2+-}\upsilon
-i\kappa z\,\bar\upsilon\Gamma_{11}\Gamma_{2+-}\upsilon
+\dots
\eea
where we have used the expansion of the $AdS$ vielbein and spin connection to $\mathcal O(z)$
\begin{equation}
\label{eq:AdSexpansion}
e^+\sim dz^+-zdz^-\,,\qquad e^-\sim dz^-+zdz^+\,,\qquad\omega^{2-}\sim dz^++zdz^-\,,\qquad\omega^{2+}\sim-dz^-+zdz^+\,.
\end{equation}
The quadratic fermion terms come from the spin connection inside $\mathcal D$ in (\ref{E}). We also have the Virasoro constraints $G_{++}=0=G_{--}$ where $G_{ij}=E_i{}^AE_j{}^B\eta_{AB}$ is the induced metric on the worldsheet. Using the fact that
\begin{equation}
E_i{}^A=e_i{}^A+i\bar\Theta\Gamma^A\mathcal D_i\Theta+\mathcal O(\Theta^4)
\end{equation}
we find
\begin{eqnarray}
\label{eq:virasoro}
0&=&\frac{1}{4\kappa}G_{++}=-\partial_+z^--\kappa z+\frac{i}{4}\bar\upsilon\Gamma_{2+-}\upsilon+\ldots\\
0&=&\frac{1}{4\kappa}G_{--}=-\partial_-z^++\kappa z+\frac{i}{4}\bar\upsilon\Gamma_{2+-}\upsilon+\ldots\nonumber
\end{eqnarray}
where we have dropped all terms of dimension greater than one. Again the fermion terms come from the $AdS$ spin connection (\ref{eq:AdSexpansion}). Using (\ref{eq:virasoro}) to solve for $\partial_\pm z^\mp$ allows us to write $\mathcal L_z$ as
\bea
\mathcal{L}_z=
-2\kappa^2z^2
-i\kappa z\,\bar\upsilon\Gamma_{11}\Gamma_{2+-}\upsilon
-\frac{1}{8}(\bar\upsilon\Gamma_{2+-}\upsilon)^2
+\ldots
\eea
Since the kinetic term for the massive boson is $\frac{1}{2}\partial_+z\partial_-z$ we see that $z$ has mass $2\kappa$. Finally, integrating out the massive boson $z$ gives
\bea\label{eq:Lz-final}
\mathcal{L}_z=-\frac{1}{8}(\bar\upsilon\Gamma_{11}\Gamma_{2+-}\upsilon)^2-\frac{1}{8}(\bar\upsilon\Gamma_{2+-}\upsilon)^2\,.
\eea
As it turns out, once we impose the kappa symmetry gauge-fixing this will completely vanish. Thus, in hindsight we could have put the massive bosons and fermions in the action to zero directly but only if we fix the kappa symmetry in a certain way.

\subsection{z-independent part}
In the previous section we took care of all terms in the low-energy effective action which involved the $AdS$-coordinates. The remaining terms are obtained by simply setting $(z^\pm,z)$ to zero (recall that we are also setting the massive (coset) fermions to zero). From (\ref{eq:L0}) the purely bosonic part of the Lagrangian is simply
\begin{equation}
\mathcal L^{(0)}=\frac{1}{2}e_+{}^{\hat a}e_-{}^{\hat a}+\frac{1}{2}e_+{}^{a'}e_-{}^{a'}+\frac{1}{2}\partial_+y\partial_-y\,,
\end{equation}
where $e^{\hat a}$ and $e^{a'}$ are the vielbeins of the first and second $S^3$ respectively and $y$ is the $S^1$ coordinate.

Using (\ref{eq:L2}), (\ref{E}) and, that to leading order, $e_+{}^+\sim e_-{}^-\sim\omega_+{}^{2-}\sim-\omega_-{}^{2+}\sim\kappa$ the terms quadratic in fermions become
\begin{eqnarray}
\mathcal L^{(2)}&=&i\bar\upsilon\Gamma_+P(\partial_--\frac{1}{4}\omega_-^{A'B'}\Gamma_{A'B'})\upsilon
+i\bar\upsilon\Gamma_-P(\partial_+-\frac{1}{4}\omega_+^{A'B'}\Gamma_{A'B'})\upsilon
\nonumber\\
&&{}
+\frac{i}{2}e_+{}^{A'}\,\bar\upsilon\Gamma_{A'}\Gamma_{2+}P\upsilon
-\frac{i}{2}e_-{}^{A'}\,\bar\upsilon\Gamma_{A'}\Gamma_{2-}P\upsilon
+\ldots
\label{eq:action-semi-final}
\end{eqnarray}
where the ellipsis denotes terms of mass dimension higher than two which are to be dropped in the low-energy effective action. Here $A'=(\hat a,a')$ runs over the indices of the two $S^3$'s (note that the $S^1$-coordinate $y$ decouples due to the projection condition $\upsilon=(1-\mathcal P)\upsilon$ with $\mathcal P$ given in (\ref{calP})). Also note that the coupling of the vielbeins to the fermions breaks the 2d Lorentz invariance. This is in contrast to the $AdS_5\times S^5$ and $AdS_4\times\mathbbm{CP}^3$ case where the low-energy effective action is Lorentz invariant.

The kappa-symmetry projector $P$ is defined in (\ref{eq:kappaP}). A natural choice of kappa-symmetry gauge is\footnote{The more standard gauge $\Gamma^+\Theta=0$ is clearly not a good choice here since the fermion kinetic operator degenerates in this case.}
\begin{equation}
\label{eq:kappa-gauge}
\upsilon=P\upsilon=\frac12(1+\Gamma^{01}\Gamma_{11})\upsilon\,.
\end{equation}
This gauge has the additional benefit that $\mathcal{L}_z$, the the terms in the Lagrangian resulting from integrating out $z$, in (\ref{eq:Lz-final}) vanishes. We are left with the low-energy effective action
\bea
\mathcal{L}&=&\mathcal L^{(0)}+i\bar\upsilon\Gamma_+\partial_-\upsilon+i\bar\upsilon\Gamma_-\partial_+\upsilon
+\frac{i}{2}e_+{}^{A'}\,\bar\upsilon\Gamma_{A'2+}\upsilon
-\frac{i}{2}e_-{}^{A'}\,\bar\upsilon\Gamma_{A'2-}\upsilon
\nonumber\\
&&{}
-\frac{i}{4}\omega_-{}^{A'B'}\,\bar\upsilon\Gamma_{A'B'+}\upsilon
-\frac{i}{4}\omega_+{}^{A'B'}\,\bar\upsilon\Gamma_{A'B'-}\upsilon
+\mathcal L^{(4)}
\label{eq:quadratic}
\eea 
where $\mathcal{L}^{(4)}$ denotes the $\upsilon^4$-terms arising from (\ref{eq:L4}). We will now show that in fact these terms give no contribution to the low-energy effective action in the present case.

The third term in (\ref{eq:L4}) is easily seen to give no contribution at dimension two due to the projection condition $\upsilon=(1-\mathcal P)\upsilon$, the form of $S$ in (\ref{E}) and the fact that $(1-\mathcal P)\Gamma_a=\Gamma_a\mathcal P$ ($a=0,1,2$). Using (\ref{eq:quartic-matrices}), the fact that the only contribution from $\mathcal D$ comes from the $AdS$ spin connection and the kappa-gauge choice $\upsilon=P\upsilon$ one can show that the second term in (\ref{eq:L4}) also gives no contribution. The remaining terms become, after some simplification,
\bea
\mathcal L^{(4)}&=&
\frac{1}{4}\bar\upsilon\Gamma^{A'}{}_{2-}\upsilon\,\bar\upsilon\Gamma_{A'2+}\upsilon 
+\frac{1}{48}\bar\upsilon\Gamma_+{}^{CD}\upsilon\,\big(3\bar\upsilon\Gamma_-(1-\Gamma_{11})U_{CD}\upsilon-2\bar\upsilon\Gamma_C(1+\Gamma_{11})U_{D-}\upsilon\big) 
\nonumber\\
&&{}
+\frac{1}{48}\bar\upsilon\Gamma_-{}^{CD}\upsilon\,\big(3\bar\upsilon\Gamma_+(1+\Gamma_{11})U_{CD}\upsilon-2\bar\upsilon\Gamma_C(1-\Gamma_{11})U_{D+}\upsilon\big)\,. 
\eea
Using the form of $U$ in (\ref{eq:quartic-matrices}) the contribution from the $U_{D\pm}$-terms is easily seen to vanish due to the kappa gauge condition (\ref{eq:kappa-gauge}). For the remaining $U$-terms only the term involving the Riemann tensor contributes due to the fact that $\upsilon=(1-\mathcal P)\upsilon$ and we are left with
\bea
\mathcal L^{(4)}&=&
\frac{1}{4}\bar\upsilon\Gamma^{A'}{}_{2-}\upsilon\,\bar\upsilon\Gamma_{A'2+}\upsilon 
-\frac{1}{16}R_{AB}{}^{CD}\,\bar\upsilon\Gamma_+{}^{AB}\upsilon\,\bar\upsilon\Gamma_{CD-}\upsilon\,.
\eea
Using the relations
\begin{equation}
\mathcal P\Gamma^{a'}(1-\mathcal P)=\frac12\sqrt{1-\alpha}\,\varepsilon^{a'b'c'}\,\Gamma^{012}\Gamma_{b'c'}(1-\mathcal P)\,,
\qquad\mathcal P\Gamma^{\hat a}(1-\mathcal P)=\frac12\sqrt{\alpha}\,\varepsilon^{\hat a\hat b\hat c}\,\Gamma^{012}\Gamma_{\hat b\hat c}(1-\mathcal P)
\end{equation}
and the form of the Riemann tensor in (\ref{eq:Riemann}) we find that the two terms cancel so that there is no contribution from the quartic fermion terms to the low-energy effective action. This is in contrast to the $AdS_4\times\mathbbm{CP}^3$ case where such terms were found to contribute \cite{Bykov:2010tv}.

\subsection{2d fermion notation}
\label{sec:2dform}
It will be useful to write the action using a 2d notation for the fermions. Before gauge-fixing we have sixteen real massless non-coset fermions $\upsilon$. Fixing the kappa-symmetry gauge (\ref{eq:kappa-gauge}) reduces these to eight real fermions. These can be combined into four two-component Majorana fermions $\Psi_I$ ($I=1,\ldots,4$) as described in detail in appendix \ref{sec:fermions}. The Lagrangian (\ref{eq:quadratic}) then becomes
\begin{eqnarray}
\lefteqn{
\mathcal{L}=\mathcal L^{(0)}
+\frac{i}{2}\bar\Psi_I\rho^i\partial_i\Psi_I
-\frac{i}{2}(\omega_i{}^{34}+\sqrt{\alpha}\,\delta_{ij}e^{j5})\,(\bar\Psi_1\rho^i\Psi_2+\bar\Psi_3\rho^i\Psi_4)
}
\nonumber\\
&&{}
-\frac{i}{2}(\omega_i{}^{53}+\sqrt{\alpha}\,\delta_{ij}e^{j4})\,(\bar\Psi_2\rho^i\Psi_4-\bar\Psi_1\rho^i\Psi_3)
-\frac{i}{2}(\omega_i{}^{45}+\sqrt{\alpha}\,\delta_{ij}e^{j3})\,(\bar\Psi_1\rho^i\Psi_4+\bar\Psi_2\rho^i\Psi_3)
\nonumber\\
&&{}
+\frac{i}{2}(\omega_i{}^{78}+\sqrt{1-\alpha}\,\delta_{ij}e^{j6})\,(\bar\Psi_1\rho^i\Psi_4-\bar\Psi_2\rho^i\Psi_3)
+\frac{i}{2}(\omega_i{}^{86}+\sqrt{1-\alpha}\,\delta_{ij}e^{j7})\,(\bar\Psi_2\rho^i\Psi_4+\bar\Psi_1\rho^i\Psi_3)
\nonumber\\
&&{}
+\frac{i}{2}(\omega_i{}^{67}+\sqrt{1-\alpha}\,\delta_{ij}e^{j8})\,(\bar\Psi_3\rho^i\Psi_4-\bar\Psi_1\rho^i\Psi_2)\,.
\end{eqnarray}
Worldsheet indices ($i,j,\ldots$) are raised and lowered with the worldsheet metric $\eta_{ij}$ with $\eta_{+-}=2$. Note that the explicit appearance of $\delta_{ij}$ ($\delta_{++}=\delta_{--}=2$) breaks the 2d Lorentz-invariance of the action. This somewhat complicated Lagrangian can be written in a much simpler and more illuminating form by combining the four spinors into a $2\times2$ matrix
\bea 
\Psi=\sigma^I\Psi_I\,,\quad \sigma^I=\big(\sigma^1,\,\sigma^2,\,\sigma^3,\,i\mathbbm 1\big)\,,\qquad\left(\bar\Psi=\Psi^\dagger\rho^0=\bar\sigma^I\bar\Psi_I\,,\quad\bar\sigma^I=\big(\sigma^1,\,\sigma^2,\,\sigma^3,\,-i\mathbbm 1\big)\right)\,.
\eea 
The Lagrangian then becomes
\bea 
\mathcal L&=&\frac12e_i{}^{\hat a}e^{i\hat a}+\frac12e_i{}^{a'}e^{ia'}+\frac12\partial_iy\partial^iy
+\frac{i}{4}\mathrm{tr}\big(\bar\Psi\rho^i\partial_i\Psi\big)
-\frac{1}{16}(\varepsilon^{\hat a\hat b\hat c}\omega_i{}^{\hat b\hat c}+2\sqrt{\alpha}\,\delta_{ij}e^{j\hat a})\,\mathrm{tr}\big(\bar\Psi\rho^i\Psi\sigma^{\hat a}\big)
\nonumber\\
&&{}
+\frac{1}{16}(\varepsilon^{a'b'c'}\omega_i{}^{b'c'}+2\sqrt{1-\alpha}\,\delta_{ij}e^{ja'})\mathrm{tr}\big(\bar\Psi\rho^i\sigma^{a'}\Psi\big)\,,
\label{eq:LPsi}
\eea 
where $\sigma^{\hat a}=(\sigma^1,\sigma^2,\sigma^3)$ and similarly for $\sigma^{a'}$. This Lagrangian describes two $O(4)$ sigma models which are coupled through their interactions with the fermions. In addition there is a completely decoupled scalar $y$ coming from the $S^1$. The action is invariant under the two $SO(3)\sim SU(2)$ which correspond to rotations in the first and second $S^3$ factor of $AdS_3\times S^3\times S^3\times S^1$. The fermions transform as
\begin{equation}
\Psi\rightarrow U^\dagger\Psi V\qquad(\bar\Psi\rightarrow V^\dagger\bar\Psi U)\qquad\mbox{with}\qquad U\in SU(2)_1\,,\quad V\in SU(2)_2\,.
\end{equation}
In fact (\ref{eq:LPsi}) is invariant under the full isometry group of the two $S^3$ i.e. $SO(4)\times SO(4)$. This can for example be seen by verifying explicitly the invariance under the (appropriately restricted) superisometry transformations given in \cite{Sundin:2012gc}. It also follows from the Lax connection construction in the next section. Unlike the $AdS_5\times S^5$ and $AdS_4\times\mathbbm{CP}^3$ case we have not found a form of the Lagrangian that makes the full $SO(4)\times SO(4)$ symmetry manifest.

As a side note, the fermion terms in the action can be written in a more compact form as
\begin{equation}
\frac{i}{4}\mathrm{tr}\big(\bar\Psi\rho^iD_i\Psi\big)\,,
\end{equation}
where the generalized ''covariant'' derivative is defined as
\begin{equation}
\label{eq:Di}
D_i\Psi=\partial_i\Psi
+\frac{i}{4}(\varepsilon^{\hat a\hat b\hat c}\omega_i{}^{\hat b\hat c}+2\sqrt{\alpha}\,\delta_{ij}e^{j\hat a})\Psi\sigma^{\hat a}
-\frac{i}{4}(\varepsilon^{a'b'c'}\omega_i{}^{b'c'}+2\sqrt{1-\alpha}\,\delta_{ij}e^{ja'})\sigma^{a'}\Psi
\end{equation}
or, equivalently,
\begin{equation}
D_i\Psi=\partial_i\Psi-\frac14\omega'_i{}^{IJ}\sigma_{IJ}\Psi+\frac14\hat\omega_i{}^{IJ}\Psi\sigma_{IJ}\qquad\sigma^{IJ}=\bar\sigma^{[I}\sigma^{J]}
\end{equation}
with
\begin{equation}
\omega'_i{}^{a'b'}=\omega_i{}^{a'b'}\,,\quad
\omega'_i{}^{a'4}=\sqrt{1-\alpha}\,\delta_{ij}e^{ja'}\,,\quad
\hat\omega_i{}^{\hat a\hat b}=\omega_i{}^{\hat a\hat b}\,,\quad
\hat\omega_i{}^{\hat a4}=\sqrt\alpha\,\delta_{ij}e^{j\hat a}\,.
\end{equation}
Note however that $D_i\Psi$ is \emph{not} 2d Lorentz-covariant.

So far what we have said refers to the $AdS_3\times S^3\times S^3\times S^1$ case. The $AdS_3\times S^3\times T^4$ case is however easily obtained by taking the limit $\alpha\rightarrow1$ (or, equivalently, $\alpha\rightarrow0$) so that $e_i{}^{a'}\rightarrow \partial_iy^{a'}$ and $\omega_i{}^{a'b'}\rightarrow0$. This leads to an $O(4)$ sigma model coupled to four Majorana fermions together with four decoupled scalars from the $T^4$.

\section{Classical integrability}
\label{sec:integrability}
Since the low-energy effective action has no quartic fermion terms its Lax connection can be obtained from \cite{Sundin:2012gc} by a suitable truncation. In that paper a Lax connection was written for the complete $AdS_3\times S^3\times S^3\times S^1$ superstring with 32 fermions, up to quadratic order in fermions (see \cite{Sorokin:2010wn,Sorokin:2011rr,Cagnazzo:2011at} for similar Lax connections for $AdS_4\times\mathbbm{CP}^3$ and $AdS_2\times S^2\times T^6$). Normally this Lax connection would have zero curvature, i.e.
\begin{equation}
\label{eq:dL}
dL-L\wedge L=0\qquad\mbox{or}\qquad \varepsilon^{ij}(\partial_iL_j+L_iL_j)=0\,,
\end{equation}
only modulo $\Theta^4$-terms. Additional terms would then be needed in the Lax connection at order $\Theta^4$ to cancel these terms. However, in the present case the low-energy effective action actually terminates at the quadratic order in fermions. Since we expect this model to be integrable we should find that the $\upsilon^4$-terms coming from $L\wedge L$ actually cancel in this case. We will now show that this is precisely what happens.

The Lax connection splits into two pieces\footnote{We drop the additional $S^1$ boson $y$ since it decouples completely. It can of course be trivially included in the Lax connection.}
\begin{equation}
L=\hat L+L'\,,
\end{equation}
coming from the two $S^3$'s respectively. In terms of the components of the Maurer-Cartan form $K$ on $S^3$ satisfying (see \cite{Sundin:2012gc} for more details)
\begin{equation}
\label{eq:MC}
dK=K\wedge K\quad\Rightarrow [K_{\hat a},K_{\hat b}]=\nabla_{\hat a}K_{\hat b}\,,\quad[K_{\hat c},\nabla_{\hat a}K_{\hat b}]=-2\alpha\delta_{\hat c[\hat a}K_{\hat b]}\,,
\end{equation}
and similarly for $K_{a'}$ with $\alpha\rightarrow1-\alpha$, these Lax connection pieces are given by
\begin{eqnarray}
\hat L_i&=&(\alpha_1\delta_i^j+\alpha_2\eta_{ik}\varepsilon^{kj})e_j{}^{\hat a}\,K_{\hat a}
-\frac{\alpha_2}{8}\sqrt\alpha\,\varepsilon_{ij}\delta^{jk}\mathrm{tr}\big(\bar\Psi\rho_k\Psi\sigma^{\hat a}\big)\,K_{\hat a}
\nonumber\\
&&{}
-\frac{\alpha_2}{16}(\alpha_2\eta_{ij}+(1+\alpha_1)\varepsilon_{ij})\varepsilon^{\hat a\hat b\hat c}\mathrm{tr}\big(\bar\Psi\rho^j\Psi\sigma_{\hat c}\big)\,\nabla_{\hat a}K_{\hat b}
\\
&&{}\nonumber\\
L'_i&=&(\alpha_1\delta_i^j+\alpha_2\eta_{ik}\varepsilon^{kj})e_j{}^{a'}\,K_{a'}
+\frac{\alpha_2}{8}\sqrt{1-\alpha}\,\varepsilon_{ij}\delta^{jk}\mathrm{tr}\big(\bar\Psi\rho_k\sigma^{a'}\Psi\big)\,K_{a'}
\nonumber\\
&&{}
+\frac{\alpha_2}{16}(\alpha_2\eta_{ij}+(1+\alpha_1)\varepsilon_{ij})\varepsilon^{a'b'c'}\mathrm{tr}\big(\bar\Psi\rho^j\sigma_{c'}\Psi\big)\,\nabla_{a'}K_{b'}\,.
\end{eqnarray}
These two pieces obviously commute with each other since they are constructed using generators from different algebras. It is also worth noting that due to the explicit appearance of $\delta_{ij}$ $L_i$ is not covariant under 2d Lorentz-transformations. This is to be expected since the low-energy effective action lacks this symmetry as we have seen. The parameters $\alpha_1$ and $\alpha_2$ are related by the equation
\begin{equation}
\label{eq:spectral}
\alpha_2^2=2\alpha_1+\alpha_1^2
\end{equation}
and can therefore be expressed in terms of a single (spectral) parameter. Let us now show that the curvature (\ref{eq:dL}) of $\hat L_i$ indeed vanishes on-shell. The same is true for $L_i'$ by an essentially identical calculation. Computing the second term in (\ref{eq:dL}) we find using (\ref{eq:spectral})
\begin{eqnarray}
\varepsilon^{ij}\hat L_i\hat L_j
&=&
-\alpha_1\varepsilon^{ij}e_i{}^{\hat a}e_j{}^{\hat b}\,[K_{\hat a},K_{\hat b}]
+\frac{\alpha_2}{16}e_i{}^{\hat d}
\Big(
2\sqrt\alpha\,(\alpha_1\delta^i_k-\alpha_2\varepsilon^{ij}\eta_{jk})\delta^{kl}\mathrm{tr}\big(\bar\Psi\rho_l\Psi\sigma^{\hat a}\big)\,[K_{\hat a},K_{\hat d}]
\nonumber\\
&&{}
+(\alpha_1\delta_k^i+\alpha_2\varepsilon^{ij}\eta_{jk})\varepsilon^{\hat a\hat b\hat c}\mathrm{tr}\big(\bar\Psi\rho^k\Psi\sigma_{\hat c}\big)\,[K_{\hat d},\nabla_{\hat a}K_{\hat b}]
\Big)
\nonumber\\
&&{}
+\frac{\alpha_2^2}{256}
\Big(
\varepsilon^{ij}\mathrm{tr}\big(\bar\Psi\rho_i\Psi\sigma^{\hat a}\big)\mathrm{tr}\big(\bar\Psi\rho_j\Psi\sigma^{\hat b}\big)
(
2\alpha[K_{\hat a},K_{\hat b}]
+\frac12\varepsilon^{\hat a\hat c\hat d}\varepsilon^{\hat b\hat e\hat f}[\nabla_{\hat e}K_{\hat f},\nabla_{\hat c}K_{\hat d}]
)
\nonumber\\
&&{}
-2\sqrt\alpha\,\delta^{ik}(\alpha_2\eta_{kj}+(1+\alpha_1)\varepsilon_{kj})
\mathrm{tr}\big(\bar\Psi\rho_i\Psi\sigma^{\hat c}\big)\mathrm{tr}\big(\bar\Psi\rho^j\Psi\sigma_{\hat d}\big)
\varepsilon^{\hat a\hat b\hat d}[K_{\hat c},\nabla_{\hat a}K_{\hat b}]
\Big)\,.
\end{eqnarray}
Using the relations (\ref{eq:MC}) it is easy to see that the terms quartic in $\Psi$ indeed vanish as advertised earlier. Using these relations and the fact that $\partial_iK_{\hat b}=e_i{}^{\hat a}\nabla_{\hat a}K_{\hat b}+\omega_{i\hat b}{}^{\hat c}K_{\hat c}$ we get for the curvature of $\hat L_i$
\begin{eqnarray}
\varepsilon^{ij}(\partial_i\hat L_j+\hat L_i\hat L_j)
&=&
\alpha_2\big(\nabla^ie_i{}^{\hat a}-\frac{\sqrt\alpha}{4}\mathrm{tr}\big(\bar\Psi\rho^i\delta_{ij}D^j\Psi\sigma^{\hat a}\big)\big)\,K_{\hat a}
\nonumber\\
&&{}
+i\frac{\alpha_2}{8}\mathrm{tr}\big(\bar\Psi((1+\alpha_1)+\alpha_2\rho^3)\rho^iD_i\Psi\sigma^{\hat a\hat b}\big)\,\nabla_{\hat a}K_{\hat b}\,,
\label{eq:dhatL}
\end{eqnarray}
where $\rho^3=\rho^0\rho^1$. The first term is the equation of motion for $y^{\hat m}$ following from the action (\ref{eq:LPsi}) (modulo a term proportional to the fermionic equation of motion) and the second term is proportional to the fermionic equation of motion. Note however that the $e^{a'}$ and $\omega^{a'b'}$-terms inside $D_i$ in (\ref{eq:Di}) don't contribute here due to the fact that
\begin{equation}
\mathrm{tr}\big(\bar\Psi\rho^i\sigma^{a'}\Psi\sigma^{\hat a}\big)=0\,.
\end{equation}
It is clear that the curvature of $\hat L_i$ (\ref{eq:dhatL}) vanishes on-shell. It is also clear that the flatness of $\hat L_i$ by itself does not imply all the equations of motion, only the equations for $y^{\hat m}$ and part of the fermionic equation due to the missing contributions involving $e^{a'}$ and $\omega^{a'b'}$ as mentioned above. However, together the flatness of $\hat L_i$ and $L_i'$ imply all the equations of motion of (\ref{eq:LPsi}) demonstrating the classical integrability of the model.

\section{Conclusions}
In this paper we have derived the Lagrangian that captures the low-energy dynamics of fluctuations of the $AdS_3\times S^3\times M_4$ string around the GKP vacuum \cite{Gubser:2002tv}. The starting point of our analysis was the GS action up to quartic order in fermions, recently derived in \cite{Wulff:2013kga}. The classical GKP solution is a spinning string in $AdS_3$ with a fluctuation spectrum of both massive and massless modes. While the full fluctuation Lagrangian is very involved, it simplifies drastically in the low-energy limit where only the massless modes contribute. We have found that the resulting theory consists of two $O(4)$ sigma models coupled through their coupling to four Majorana fermions. In addition there is a free boson coming from the $S^1$. In contrast to earlier examples in $AdS_5\times S^5$ and $AdS_4\times \mathbbm{CP}^3$ we find that the model is not 2d Lorentz invariant. Furthermore, a curious fact is that the quartic fermion terms completely drop out due to a delicate cancellation. This is somewhat unexpected, since, at least for $\alpha=\frac{1}{2}$, the $AdS_4\times \mathbbm{CP}^3$ and $AdS_3\times S^3\times S^3\times S^1$ strings share many similar features, see \cite{Sundin:2012gc} for some examples. The special case where one $S^3$ blows up (i.e. $\alpha\rightarrow 0,1$) describes the $AdS_3\times S^3\times T^4$ GKP string and the low-energy effective action reduces to a single $O(4)$ sigma model coupled to four Majorana fermions with four free bosons coming from the $T^4$. 

There are several interesting possible extensions of this work. It would be very interesting to perform a similar analysis as \cite{Basso:2012bw,Basso:2013pxa,Fioravanti:2013eia} and try to find the Bethe ansatz and exact S-matrix for this model. Since the full $AdS_3\times S^3\times S^3\times S^1$ string is believed to be quantum integrable it is natural to expect that the low-energy string should inherit this integrability. For example, as in \cite{Basso:2013pxa}, one can derive the S-matrix for the low-energy excitations and match the resulting asymptotic equations with the low-energy part of the full set of conjectured equations in \cite{Borsato:2013hoa,Borsato:2013qpa}.

It would also be interesting to derive the low-energy limit of the GKP string for the case of $AdS_3\times S^3\times S^3\times S^1$ with mixed RR and NSNS flux. Superficially this case is considerably more complicated but recent S-matrix calculations \cite{Hoare:2013pma,Hoare:2013ida} have shown that this case is very similar to the pure RR flux case. It would be interesting to understand what happens in the limit of pure NSNS flux since one can then connect to the RNS description of the string.

\section*{Acknowledgments}
It's a pleasure to thank M. Abbott and A. Rej for discussions and comments. P.S acknowledges the support of a Claude Leon postdoctoral grant. The research of LW is supported in part by NSF grant PHY-0906222.

\appendix

\section{2d spinors and gamma-matrices}\label{sec:fermions}
Starting from the fermions $\upsilon$ with eight real components (after fixing the kappa symmetry (\ref{eq:kappa-gauge})) we define one-component fermions by the following projections
\begin{equation}
\psi_{\pm\pm\pm}=\frac12(1\pm\Gamma^{01})\frac12(1\pm i\Gamma^{34})\frac12(1\pm i\Gamma^{67})\upsilon\,.
\end{equation}
In the space of $\upsilon$ the gamma-matrices are effectively eight-dimensional and we take a realization such that
\begin{eqnarray}
C\sim-i\sigma^2\otimes\sigma^2\otimes\sigma^2\,,\quad\Gamma_\pm\sim i\sigma^2(\mathbbm1\pm\sigma^3)\otimes\mathbbm1\otimes\mathbbm1\,,\quad
%\nonumber\\
\Gamma^{45}\sim\mathbbm1\otimes i\sigma^1\otimes\mathbbm1\,,\quad\Gamma^{78}\sim\mathbbm1\otimes\mathbbm1\otimes i\sigma^1\,,
%C\Gamma_\pm=(\mathbbm1\pm\sigma^3)\otimes\sigma^2\otimes\sigma^2
\end{eqnarray}
where $C$ is the charge-conjugation matrix. As defined the spinors $\psi$ are not real but satisfy
\begin{equation}
\psi_{\pm+-}^\dagger=\psi_{\pm-+}\,,\qquad\psi_{\pm++}^\dagger=-\psi_{\pm--}\,,
\end{equation}
as follows from the Majorana condition on $\upsilon$. We can define real spinors as
\bea
\begin{array}{ll}
\psi_1=\sqrt2(\psi_{---}-\psi_{-++})\,,&\psi_2=i\sqrt2(\psi_{---}+\psi_{-++})\,,\\  
\psi_3=-\sqrt2(\psi_{-+-}+\psi_{--+})\,,&\psi_4=i\sqrt2(\psi_{-+-}-\psi_{--+})\,,\\  
{}\\
\chi_1=\sqrt2(\psi_{+--}-\psi_{+++})\,,&\chi_2=i\sqrt2(\psi_{+--}+\psi_{+++})\,,\\  
\chi_3=-\sqrt2(\psi_{++-}+\psi_{+-+})\,,&\chi_4=i\sqrt2(\psi_{++-}-\psi_{+-+})\,.
\end{array}
\eea
These can be combined into four 2d Majorana spinors
\begin{equation}
\Psi_I=\left(\begin{array}{c}
	\psi_I\\
	\chi_I
\end{array}\right)\qquad(I=1,\ldots,4)\,.
\end{equation}
We take the 2d gamma-matrices to be 
\begin{equation}
\rho^0=\sigma^2\,,\quad\rho^1=i\sigma^1\,,\qquad\rho^\pm=\frac{1}{2}(\rho^0\pm\rho^1)
\end{equation}
and the conjugate spinor is defined as $\bar\Psi=\Psi^\dagger\rho^0$.

%%%%%%%%%%%%%%%%%%%%


\begin{thebibliography}{50}

%\cite{Cagnazzo:2012se}
\bibitem{Cagnazzo:2012se}
  A.~Cagnazzo and K.~Zarembo,
  ``B-field in AdS(3)/CFT(2) Correspondence and Integrability,''
  JHEP {\bf 1211} (2012) 133
   [Erratum-ibid.\  {\bf 1304} (2013) 003]
  [arXiv:1209.4049 [hep-th]].
  %%CITATION = ARXIV:1209.4049;%%
  %14 citations counted in INSPIRE as of 25 Jun 2013

%\cite{Babichenko:2009dk}
\bibitem{Babichenko:2009dk}
  A.~Babichenko, B.~Stefanski, Jr. and K.~Zarembo,
  ``Integrability and the AdS(3)/CFT(2) correspondence,''
  JHEP {\bf 1003} (2010) 058
  [arXiv:0912.1723 [hep-th]].
  %%CITATION = ARXIV:0912.1723;%%
  %62 citations counted in INSPIRE as of 25 Jun 2013

%\cite{Beisert:2010jr}
\bibitem{Beisert:2010jr}
  N.~Beisert, C.~Ahn, L.~F.~Alday, Z.~Bajnok, J.~M.~Drummond, L.~Freyhult, N.~Gromov and R.~A.~Janik {\it et al.},
  ``Review of AdS/CFT Integrability: An Overview,''
  Lett.\ Math.\ Phys.\  {\bf 99} (2012) 3
  [arXiv:1012.3982 [hep-th]].
  %%CITATION = ARXIV:1012.3982;%%
  %258 citations counted in INSPIRE as of 25 Jun 2013

%\cite{OhlssonSax:2011ms}
\bibitem{OhlssonSax:2011ms} 
  O.~Ohlsson Sax and B.~Stefanski, Jr.,
  ``Integrability, spin-chains and the AdS3/CFT2 correspondence,''
  JHEP {\bf 1108}, 029 (2011)
  [arXiv:1106.2558 [hep-th]].
  %%CITATION = ARXIV:1106.2558;%%
  %22 citations counted in INSPIRE as of 25 Jun 2013

%\cite{Ahn:2012hw}
\bibitem{Ahn:2012hw} 
  C.~Ahn and D.~Bombardelli,
  ``Exact S-matrices for $AdS_3/CFT_2$,''
  arXiv:1211.4512 [hep-th].
  %%CITATION = ARXIV:1211.4512;%%
  %13 citations counted in INSPIRE as of 25 Jun 2013

%\cite{Rughoonauth:2012qd}
\bibitem{Rughoonauth:2012qd}
  N.~Rughoonauth, P.~Sundin and L.~Wulff,
  ``Near BMN dynamics of the AdS(3) x S(3) x S(3) x S(1) superstring,''
  JHEP {\bf 1207} (2012) 159
  [arXiv:1204.4742 [hep-th]].
  %%CITATION = ARXIV:1204.4742;%%
  %18 citations counted in INSPIRE as of 25 Jun 2013
  
%\cite{Borsato:2012ud}
\bibitem{Borsato:2012ud} 
  R.~Borsato, O.~Ohlsson Sax and A.~Sfondrini,
  ``A dynamic $su(1|1)^2$ S-matrix for AdS3/CFT2,''
  JHEP {\bf 1304}, 113 (2013)
  [arXiv:1211.5119 [hep-th]].
  %%CITATION = ARXIV:1211.5119;%%
  %12 citations counted in INSPIRE as of 25 Jun 2013

%\cite{Borsato:2012ss}
\bibitem{Borsato:2012ss} 
  R.~Borsato, O.~Ohlsson Sax and A.~Sfondrini,
  ``All-loop Bethe ansatz equations for AdS3/CFT2,''
  JHEP {\bf 1304}, 116 (2013)
  [arXiv:1212.0505 [hep-th]].
  %%CITATION = ARXIV:1212.0505;%%
  %8 citations counted in INSPIRE as of 25 Jun 2013

%\cite{Sundin:2013ypa}
\bibitem{Sundin:2013ypa} 
  P.~Sundin and L.~Wulff,
  ``Worldsheet scattering in AdS(3)/CFT(2),''
  arXiv:1302.5349 [hep-th].
  %%CITATION = ARXIV:1302.5349;%%
  %7 citations counted in INSPIRE as of 26 Jun 2013

%\cite{Borsato:2013qpa}
\bibitem{Borsato:2013qpa} 
  R.~Borsato, O.~Ohlsson Sax, A.~Sfondrini, B.~Stefanski, Jr. and A.~Torrielli,
  ``The all-loop integrable spin-chain for strings on AdS$_3 \times S^3 \times T^4$: the massive sector,''
  arXiv:1303.5995 [hep-th].
  %%CITATION = ARXIV:1303.5995;%%
  %4 citations counted in INSPIRE as of 25 Jun 2013

%\cite{David:2010yg}
\bibitem{David:2010yg} 
  J.~R.~David and B.~Sahoo,
  ``S-matrix for magnons in the D1-D5 system,''
  JHEP {\bf 1010}, 112 (2010)
  [arXiv:1005.0501 [hep-th]].
  %%CITATION = ARXIV:1005.0501;%%
  %21 citations counted in INSPIRE as of 28 Jun 2013

%\cite{Beccaria:2012kb}
\bibitem{Beccaria:2012kb} 
  M.~Beccaria, F.~Levkovich-Maslyuk, G.~Macorini and A.~A.~Tseytlin,
  ``Quantum corrections to spinning superstrings in $AdS_3 x S^3 x M^4$: determining the dressing phase,''
  JHEP {\bf 1304}, 006 (2013)
  [arXiv:1211.6090 [hep-th]].
  %%CITATION = ARXIV:1211.6090;%%
  %9 citations counted in INSPIRE as of 25 Jun 2013

%\cite{Beccaria:2012pm}
\bibitem{Beccaria:2012pm} 
  M.~Beccaria and G.~Macorini,
  ``Quantum corrections to short folded superstring in $AdS_3 x S^3 x M^4$,''
  JHEP {\bf 1303}, 040 (2013)
  [arXiv:1212.5672 [hep-th]].
  %%CITATION = ARXIV:1212.5672;%%
  %4 citations counted in INSPIRE as of 25 Jun 2013

%\cite{Abbott:2012dd}
\bibitem{Abbott:2012dd} 
  M.~C.~Abbott,
  ``Comment on Strings in AdS3 x S3 x S3 x S1 at One Loop,''
  JHEP {\bf 1302}, 102 (2013)
  [arXiv:1211.5587 [hep-th]].
  %%CITATION = ARXIV:1211.5587;%%
  %11 citations counted in INSPIRE as of 25 Jun 2013

%\cite{Abbott:2013mpa}
\bibitem{Abbott:2013mpa} 
  M.~C.~Abbott,
  ``The AdS3 x S3 x S3 x S1 Hernandez-Lopez Phases: a Semiclassical Derivation,''
  arXiv:1306.5106 [hep-th].
  %%CITATION = ARXIV:1306.5106;%%

%\cite{Borsato:2013hoa}
\bibitem{Borsato:2013hoa} 
  R.~Borsato, O.~Ohlsson Sax, A.~Sfondrini, B.~Stefanski and A.~Torrielli,
  ``Dressing phases of AdS3/CFT2,''
  arXiv:1306.2512 [hep-th].
  %%CITATION = ARXIV:1306.2512;%%
  %1 citations counted in INSPIRE as of 25 Jun 2013

%\cite{Sax:2012jv}
\bibitem{Sax:2012jv} 
  O.~Ohlsson Sax, B.~Stefanski, jr. and A.~Torrielli,
  ``On the massless modes of the AdS3/CFT2 integrable systems,''
  JHEP {\bf 1303}, 109 (2013)
  [arXiv:1211.1952 [hep-th]].
  %%CITATION = ARXIV:1211.1952;%%
  %12 citations counted in INSPIRE as of 25 Jun 2013

%\cite{Gubser:2002tv}
\bibitem{Gubser:2002tv} 
  S.~S.~Gubser, I.~R.~Klebanov and A.~M.~Polyakov,
  ``A Semiclassical limit of the gauge / string correspondence,''
  Nucl.\ Phys.\ B {\bf 636}, 99 (2002)
  [hep-th/0204051].
  %%CITATION = HEP-TH/0204051;%%
  %721 citations counted in INSPIRE as of 25 Jun 2013

%\cite{Forini:2012bb}
\bibitem{Forini:2012bb}
  V.~Forini, V.~G.~M.~Puletti and O.~Ohlsson Sax,
  ``Generalized cusp in $AdS_4 x CP^3$ and more one-loop results from semiclassical strings,''
  J.\ Phys.\ A {\bf 46}, 115402 (2013)
  [arXiv:1204.3302 [hep-th]].
  %%CITATION = ARXIV:1204.3302;%%
  %12 citations counted in INSPIRE as of 10 Apr 2013

%\cite{Frolov:2002av}
\bibitem{Frolov:2002av}
  S.~Frolov and A.~A.~Tseytlin,
  ``Semiclassical quantization of rotating superstring in AdS(5) x S**5,''
  JHEP {\bf 0206} (2002) 007
  [hep-th/0204226].
  %%CITATION = HEP-TH/0204226;%%
  %477 citations counted in INSPIRE as of 10 Apr 2013

%\cite{Alday:2007mf}
\bibitem{Alday:2007mf}
  L.~F.~Alday and J.~M.~Maldacena,
  ``Comments on operators with large spin,''
  JHEP {\bf 0711} (2007) 019
  [arXiv:0708.0672 [hep-th]].
  %%CITATION = ARXIV:0708.0672;%%
  %137 citations counted in INSPIRE as of 10 Apr 2013

%\cite{Bykov:2010tv}
\bibitem{Bykov:2010tv}
  D.~Bykov,
  ``The worldsheet low-energy limit of the $AdS_4 x CP^3$ superstring,''
  Nucl.\ Phys.\ B {\bf 838}, 47 (2010)
  [arXiv:1003.2199 [hep-th]].
  %%CITATION = ARXIV:1003.2199;%%
  %5 citations counted in INSPIRE as of 10 Apr 2013

%\cite{Basso:2012bw}
\bibitem{Basso:2012bw}
  B.~Basso and A.~Rej,
  ``On the integrability of two-dimensional models with U(1)xSU(N) symmetry,''
  Nucl.\ Phys.\ B {\bf 866} (2013) 337
  [arXiv:1207.0413 [hep-th]].
  %%CITATION = ARXIV:1207.0413;%%
  %4 citations counted in INSPIRE as of 25 Jun 2013

%\cite{Basso:2013pxa}
\bibitem{Basso:2013pxa}
  B.~Basso and A.~Rej,
  ``Bethe Ansaetze for GKP strings,''
  arXiv:1306.1741 [hep-th].
  %%CITATION = ARXIV:1306.1741;%%
  %1 citations counted in INSPIRE as of 25 Jun 2013

%\cite{Fioravanti:2013eia}
\bibitem{Fioravanti:2013eia} 
  D.~Fioravanti, S.~Piscaglia and M.~Rossi,
  ``On the scattering over the GKP vacuum,''
  arXiv:1306.2292 [hep-th].
  %%CITATION = ARXIV:1306.2292;%%
  
%\cite{Cvetic:1999zs}
\bibitem{Cvetic:1999zs}
  M.~Cvetic, H.~Lu, C.~N.~Pope and K.~S.~Stelle,
  ``T duality in the Green-Schwarz formalism, and the massless / massive IIA duality map,''
  Nucl.\ Phys.\ B {\bf 573} (2000) 149
  [hep-th/9907202].
  %%CITATION = HEP-TH/9907202;%%
  %97 citations counted in INSPIRE as of 04 Jun 2013

%\cite{Wulff:2013kga}
\bibitem{Wulff:2013kga}
  L.~Wulff,
  ``The type II superstring to order $\theta^4$,''
  arXiv:1304.6422 [hep-th].
  %%CITATION = ARXIV:1304.6422;%%

%\cite{Sundin:2012gc}
\bibitem{Sundin:2012gc}
  P.~Sundin and L.~Wulff,
  ``Classical integrability and quantum aspects of the AdS(3) x S(3) x S(3) x S(1) superstring,''
  JHEP {\bf 1210} (2012) 109
  [arXiv:1207.5531 [hep-th]].
  %%CITATION = ARXIV:1207.5531;%%
  %16 citations counted in INSPIRE as of 21 Jun 2013


      



%\cite{Sorokin:2010wn}
\bibitem{Sorokin:2010wn}
  D.~Sorokin and L.~Wulff,
  ``Evidence for the classical integrability of the complete $AdS_4 x CP^3$ superstring,''
  JHEP {\bf 1011} (2010) 143
  [arXiv:1009.3498 [hep-th]].
  %%CITATION = ARXIV:1009.3498;%%
  %29 citations counted in INSPIRE as of 27 Jun 2013

%\cite{Sorokin:2011rr}
\bibitem{Sorokin:2011rr} 
  D.~Sorokin, A.~Tseytlin, L.~Wulff and K.~Zarembo,
  ``Superstrings in AdS(2)xS(2)xT(6),''
  J.\ Phys.\ A {\bf 44}, 275401 (2011)
  [arXiv:1104.1793 [hep-th]].
  %%CITATION = ARXIV:1104.1793;%%
  %24 citations counted in INSPIRE as of 25 Jun 2013
  
%\cite{Cagnazzo:2011at}
\bibitem{Cagnazzo:2011at}
  A.~Cagnazzo, D.~Sorokin and L.~Wulff,
  ``More on integrable structures of superstrings in AdS(4) x CP(3) and AdS(2) x S(2) x T(6) superbackgrounds,''
  JHEP {\bf 1201} (2012) 004
  [arXiv:1111.4197 [hep-th]].
  %%CITATION = ARXIV:1111.4197;%%
  %11 citations counted in INSPIRE as of 27 Jun 2013

%\cite{Hoare:2013pma}
\bibitem{Hoare:2013pma}
  B.~Hoare and A.~A.~Tseytlin,
  ``On string theory on $AdS_3 x S^3 x T^4$ with mixed 3-form flux: tree-level S-matrix,''
  Nucl.\ Phys.\ B {\bf 873} (2013) 682
  [arXiv:1303.1037 [hep-th]].
  %%CITATION = ARXIV:1303.1037;%%
  %6 citations counted in INSPIRE as of 27 Jun 2013
  
  
%\cite{Hoare:2013ida}
\bibitem{Hoare:2013ida}
  B.~Hoare and A.~A.~Tseytlin,
  ``Massive S-matrix of $AdS_3 x S^3 x T^4$ superstring theory with mixed 3-form flux,''
  Nucl.\ Phys.\ B {\bf 873} (2013) 395
  [arXiv:1304.4099 [hep-th]].
  %%CITATION = ARXIV:1304.4099;%%
  %1 citations counted in INSPIRE as of 27 Jun 2013
  
\end{thebibliography}
\end{document}